\documentclass[epj]{webofc}
\usepackage[varg]{txfonts}   
\usepackage{graphicx}
\usepackage{caption}
\usepackage{subcaption}

\wocname{EPJ Web of Conferences}
\woctitle{ICNFP 2016}
\begin{document}
\selectlanguage{english}
\title{Haloscope searches for dark matter axions at the Center for \\Axion and Precision Physics Research}
%
%

\author{Eleni Petrakou\inst{1}\fnsep\thanks{\email{petrakou@ibs.re.kr}} for CAPP/IBS 
}

\institute{Center for Axion and Precision Physics Research, Institute for Basic Science (IBS), \\Daejeon 34141, Republic of Korea}

\abstract{%
The Center for Axion and Precision Physics Research (CAPP) was founded in 2013, with the ambition of shedding light on the strong CP problem and the proposed existence of axions. Much of CAPP's effort focuses on the direct detection of dark matter candidate axions with a series of local haloscope experiments, which endeavour to expand dramatically the coverage on the "invisible axion" mass range. The first two of them plan experimental runs during this year, tapping into ultra-low cryogenics and toroidal cavity geometries. The overall programme builds on cutting-edge technology, including developments in superconducting films, SQUID amplifiers and novel magnets. \\
This article presents the planned advancements and the status of the programme, while it can also be considered a pedagogical introduction to haloscope experiments.
}
\maketitle
\section{Motivation for axion searches}
\label{motivation}
The axion was introduced in 1978 as a consequence of the Peccei-Quinn $U(1)$ quasisymmetry\cite{1}, a popular proposed solution to the so-called strong CP problem. The strong CP problem refers to the fact that the strong force seems to conserve the CP symmetry -- although the correctness of QCD predictions requires the presence of an additional, CP- and P-violating, term of gluon interactions, unless one quark mass is zero. Adding to the puzzle, nucleons have not so far been observed to possess electric dipole moments, whose presence would imply CP- and P-violation in QCD. \\
The introduction of the, spontaneously broken, Peccei-Quinn symmetry results in the presence of a new pseudoscalar field, known as axion\cite{2}, with properties suitable for a dark matter constituent\cite{3}. (For a comprehensive overview of the Peccei-Quinn theory and the axion properties, see for example \cite{4}.) \\
\\
The Center for Axion and Precision Physics Research (CAPP)\cite{5} was founded in 2013 by the Institute for Basic Science in Daejeon, South Korea, with the goal of covering both sides of the coin of the strong CP problem: Definite measurements of the hadronic electric dipole moments and direct axion detection. Its direct detection searches, whose status is presented here, comprise of a series of experiments with the aim of excluding a significant portion of the axionic cold dark matter models' parameter space over the next years -- or, if there is an axion, finding it.

\section{Principles of haloscopes}
\label{haloscopes}
\begin{figure}[h]
\centering
\includegraphics[width=5cm]{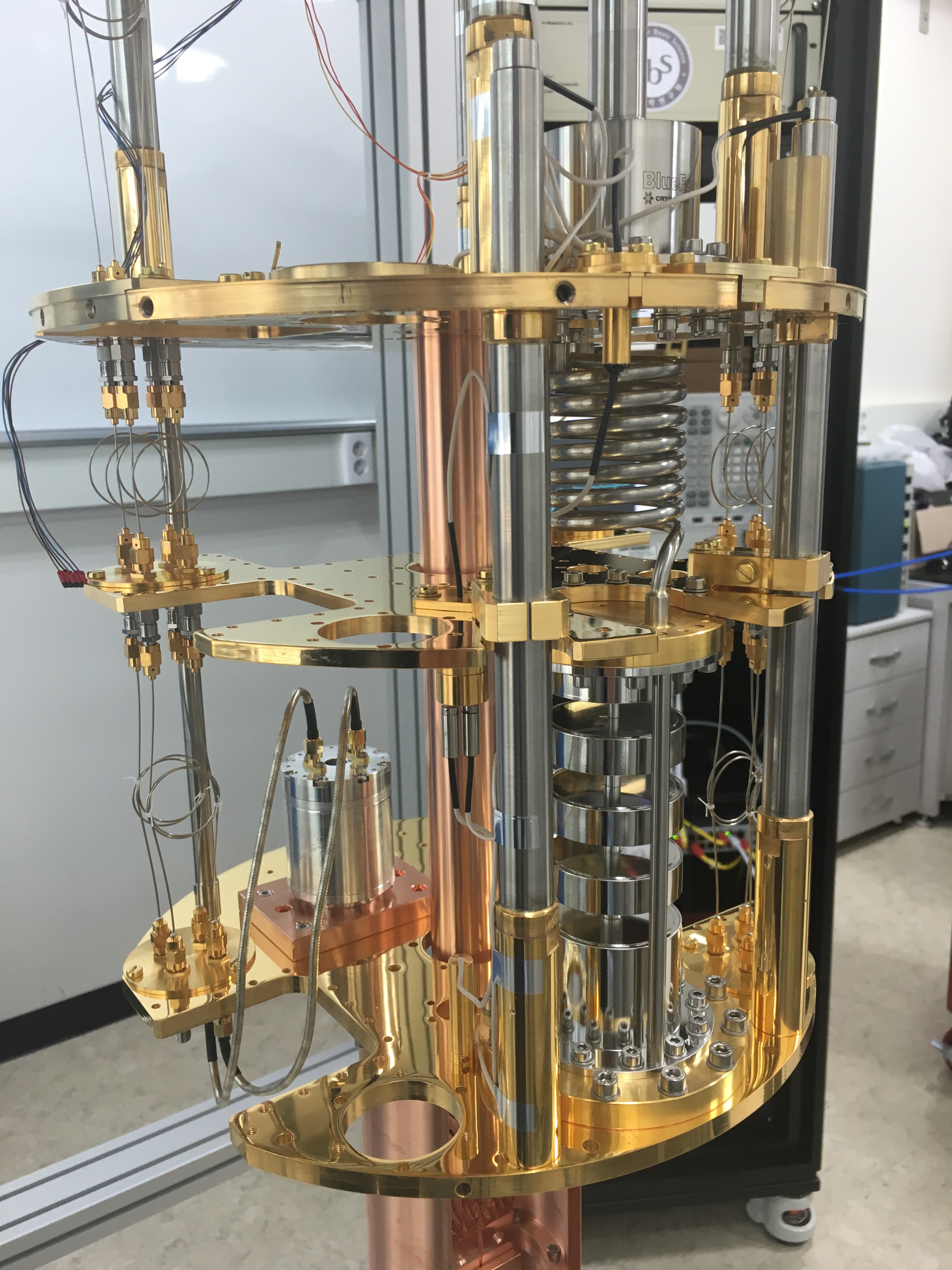}
\caption{An aluminum RF cavity (bottom left, close to image center) attached to the inside of an open dilution refrigerator and linked to the read-out electronics.}
\label{fig-1}
\end{figure}

The detection principle of haloscope experiments was first published by P. Sikivie\cite{6} in 1983, turning the thus-far considered “invisible” candidate into a detectable entity. The principle relies on the ``reverse Primakoff effect'': In the presence of a magnetic field, the axion can mix with the virtual photon at loop-level, due to the axion’s couplings to the strong sector. The end result is a real photon emitted at a frequency equal to the converted axion’s mass. As axions are predicted to be abundant if they form the galactic dark matter, the observable will consist of a continuous electromagnetic signal at the microwave frequency range. \\
Specifically, the coupling of the axion to the electromagnetic field is described by 
\[L_{\alpha\gamma\gamma}=\frac{\alpha}{\pi}\frac{g_\gamma}{f_\alpha}\,a\,\vec{E}\cdot\vec{B}, \]
where $a$ denotes the axion field, $\alpha$ is the fine structure constant, $f_\alpha$ is the axion decay constant, and $g_\gamma$ is a model-dependent coefficient of order $1$. The coupling is often expressed as $g_{\alpha\gamma\gamma} \equiv \frac{\alpha}{\pi} \frac{g_\gamma}{f_\alpha}$, and forms one of the parameters of the axion searches. \\ 
\\
Since the introduction of this principle, a number of experiments have been set up to look for axion conversions, employing strong magnetic fields. They can be broadly divided in helioscopes, which point towards the Sun for streams of solar axions, and haloscopes, which look for ambient axions from the Milky Way’s dark matter halo. \\
For realistic values of the theory, the power of the electromagnetic signal is expected to be eminently weak, with typical values lying around the range of $10^{-21}$W \cite{7}. In order to achieve sensitivity to this level of signal, haloscopes routinely operate in cryogenic temperatures for reduced noise and make use of radiofrequency cavities (“RF cavities”). The advantage of the latter is the enhancement of the signal by formation of standing waves at the cavity’s resonant frequencies. The relevant expression for the resulting signal power, inserting realistic theoretical and experimental values, is: 
\[P = 1.5\cdot10^{-24}W \left(\frac{g_{\alpha\gamma\gamma}}{10^{15}GeV}\right)^2 \left(\frac{C}{0.69}\right) \left(\frac{B}{8T}\right)^2 \left(\frac{V}{5l}\right) \left(\frac{\rho_a}{300\,MeV/cm^3}\right) \left(\frac{20\mu eV}{m_a}\right) \left(\frac{Q}{50\cdot10^3}\right),\] 
where $\rho_a$ and $m_a$ are the local density and mass of the axion, $C$ is the form factor (an expression of the overlapping of the resonant mode's electric field with the external magnetic field), $B$ is the magnetic field, and $V$ and $Q$ are the volume and loaded quality factor of the cavity. \\
The operation of RF cavities involves the appropriate choice of available electromagnetic modes, and a tuning mechanism for adjusting the resonant frequencies and so covering a range of the possible axion masses. \\
For illustration, Fig.\ref{fig-1} shows an aluminum RF cavity attached to the inside of an open dilution refrigerator at CAPP's ``RF Lab''. 

\section{Haloscope advancement at CAPP}
\label{advancement}
As seen from the last equation, the experimental settings which affect the sensitivity of a haloscope are the surrounding magnetic field, the RF cavity’s volume and quality factor, and the form factor. In addition to these, the signal-to-noise ratio improves with reduced environment and device noise temperatures. \\
CAPP prepares advancements in each of these areas with an intense in-house R\&D programme, while at the same time tapping into both local expertise and international collaboration. \\
\textbf{Cryogenic operation}\\
The haloscopes are designed to be installed inside dilution refrigerators for target operation at temperatures of a few dozen mK, and in any case well below 100mK. As of summer 2016 two BlueFors LD400 refrigerators are commissioned. In total, an array of six wet and dry dilution refrigerators will be operated in a dedicated low-vibration hall, at CAPP’s Munji Campus lab space. \\
\\
\textbf{Low-noise amplifiers}\\
HEMT amplifiers, the read-out devices of choice for weak signal, are being taken over by SQUID-based amplifiers. SQUID devices are commercially available for certain microwave frequencies, while dedicated development is carried out for the $2-10$GHz range in collaboration with the Korea Research Institute for Standards and Science \textit{(KRISS)}. \\
\\
\textbf{High field superconducting magnets}\\
High-field superconducting magnets of different strengths and bores will accommodate the various haloscopes. The first of these, to be commissioned in 2017, will provide 18T. Stronger field values will rely critically on novel high-$T_C$ cables, developed partially by the local company SuNAM. Based on this technology, the main milestone in the series of magnets will be the delivery of a 25T magnet of 10cm diameter in two years' time, as the first result of a joint project with Brookhaven National Laboratory (BNL)\cite{9}. \\
\\
\textbf{Superconducting cavity}\\
RF cavities with superconducting surfaces will decrease dramatically the power losses. While typical values for the quality factor of a copper or aluminum cavity lie at $10^4-10^5$, a fully superconducting interior would rise by several orders of magnitude (although an increase much above $10^6$ would not be beneficial to the experiments, as they would be limited by the quality factor of the axion signal). The obvious issue is maintaining superconductivity in the surrounding magnetic field. As the walls of a cylindrical cavity are parallel to the field, deposition of a thin superconducting film is expected to bring about the desired improvement. For the endcaps, a novel solution will be attempted, with etching of appropriate hole patterns allowing the magnetic flux to penetrate the surface without destroying the superconducting state of the electrons. The development of this ``vortex pinning'' technique is carried out in collaboration with the Korea Advanced Institute of Science and Technology \textit{(KAIST)}. \\
\\
\textbf{Phase-matched cavities}\\
As the ``transverse magnetic'' resonant modes chiefly used for the axion searches oscillate between the cylindrical walls, probing higher regions of the mass range requires cavities of increasingly smaller diameter. Given this restriction, one solution for increasing the effective cavity volume is the simultaneous read-out of several frequency-matched cavities of the same dimensions; as the axion oscillation length is expected to be of the order of meters, the signal from them will be coherent. A dedicated project is under way for this technique\cite{11} which has not been previously exploited in haloscope searches. \\
\\
\textbf{Toroidal RF cavities}\\
An increase in the value of the form factor can be obtained by the use of toroidal, instead of cylindrical, geometry. In addition, toroid cavities offer the advantages of reduced fringe magnetic fields and avoidance of electromagnetic mode overlapping\cite{12}.

\section{Haloscope programme and status }
\label{programme}
\begin{figure}[t]
\centering
\begin{subfigure}{.35\textwidth}
  \centering
  \includegraphics[width=3.5cm]{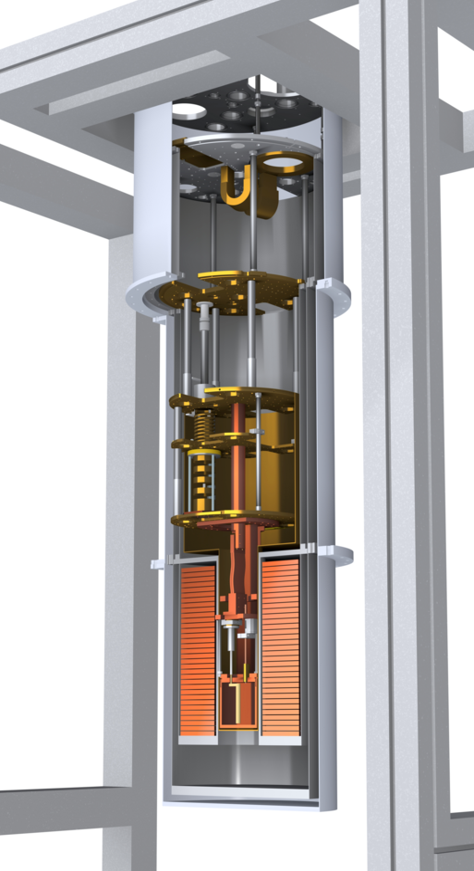}
  \label{fig:sub1}
\end{subfigure}%
\begin{subfigure}{.65\textwidth}
  \centering
  \includegraphics[width=9cm]{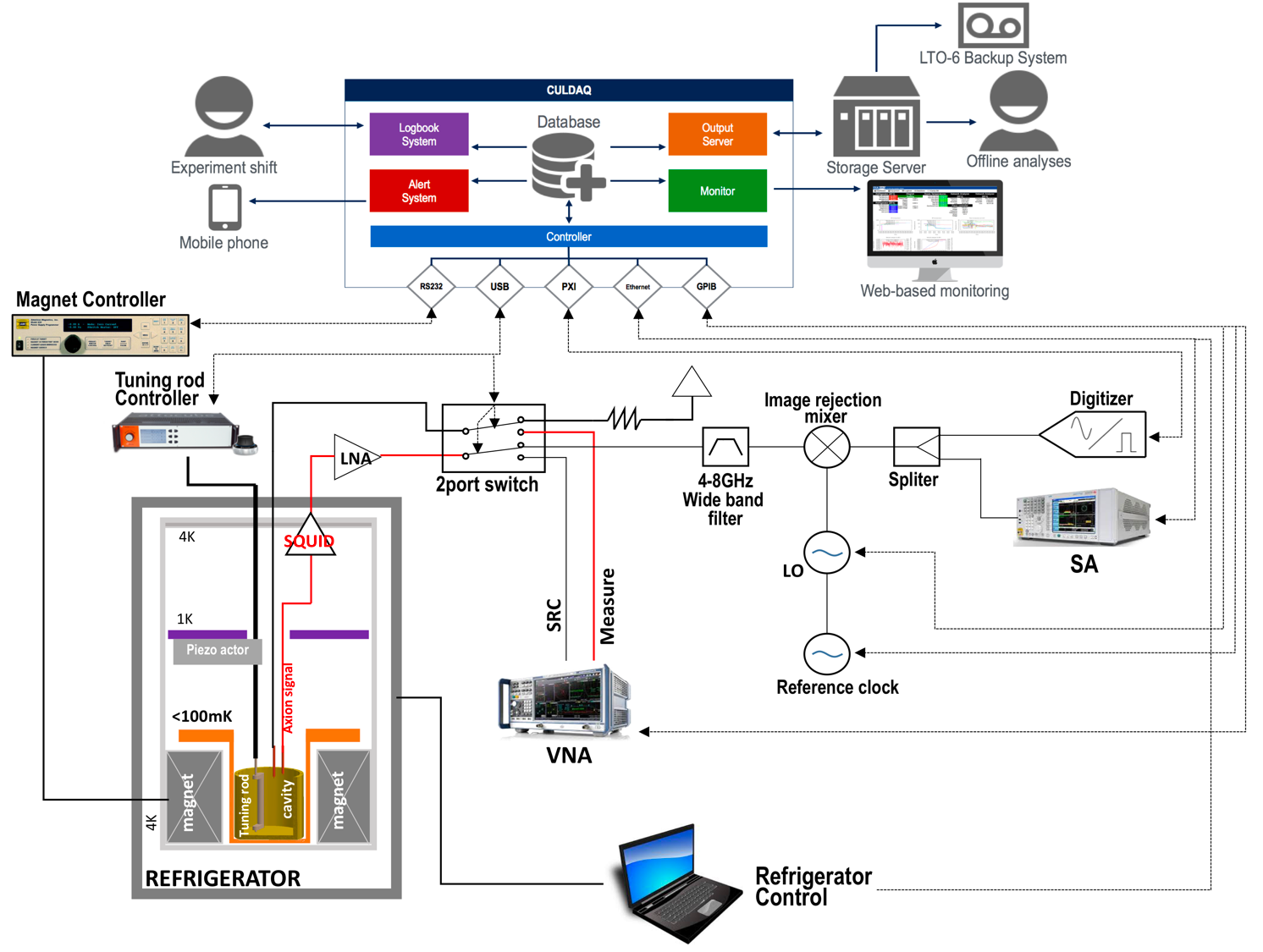}
  \label{fig:sub2}
\end{subfigure}
\caption{Left: Design of the CULTASK experiment with the BlueFors LD400 dilution regrifrator. Right: Schematic of the CULTASK electronics and data-taking system.}
\label{fig-2}
\end{figure}

\begin{figure}[hb]
\centering
\includegraphics[height=7.08cm]{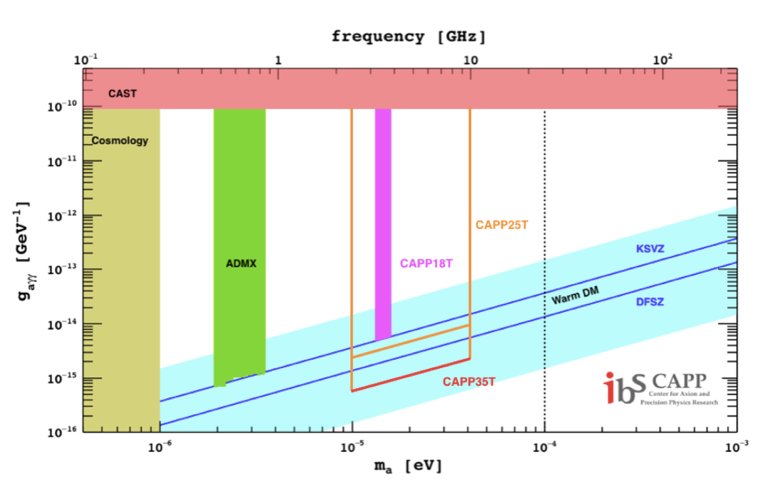}
\caption{Estimated sensitivity of CAPP haloscopes for projected magnetic fields. The blue area to the left of the dashed line corresponds to the theoretically favorable axionic cold dark matter parameters.}
\label{fig-3} 
\end{figure}

\begin{figure}[h]
\centering
\sidecaption
\includegraphics[width=6cm]{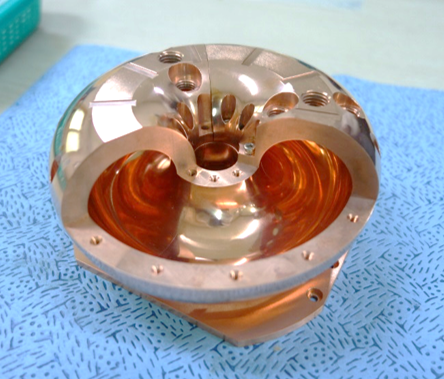}
\caption{View of the cavity used in the Toy CAPPuccino experiment.}
\label{fig-4}   
\end{figure}

\textbf{Cylindrical geometry}\\
\\
The first of the haloscope experiments employing the ``classical'' cylindrical geometry is CULTASK\cite{13} \textit{(CAPP Ultra-Low Temperature Axion Search in Korea)}. Using a copper cavity of 9cm diameter whose lowest mode corresponds to about 2GHz, it is expected to probe the $10\mu$eV axion mass vicinity and act as a precursor to the rest of the experiments. As of summer 2016 the installation, engineering runs and commissioning of the electronics chain and data acquisition system are completed, and data of reasonable sensitivity are expected within the year. \\
Fig.\ref{fig-2} shows the design of CULTASK with the cryogenic refrigerator, and the schematic of the system including the electronics and DAQ chain. \\
\\
Over the next few years stronger results are foreseen from the delivery of progressively stronger magnets and the ongoing R\&D. The commissioning of the main experiment is expected in 2018 when its key components, the 18T magnet and the custom SQUID amplifier, become available. \\
The projected sensitivity of the experiments is shown in Fig.\ref{fig-3}, in terms of the axion's mass and $g_{a\gamma\gamma}$ coupling. The estimations correspond to a signal-to-noise ratio of 5 and 100s scanning time per frequency value, for conventional quality factor cavities and estimated amplifier noise around 1K. With the addition of each further individual development, exploration of the cold dark matter parameter space will deepen, and coverage up to its higher end is expected within less than one decade. \\
\\
\\
\pagebreak
\pagebreak
\\
\textbf{Toroidal geometry}\\
The lower end of the axion mass range can be covered effectively by toroid cavities and magnets. A small copper toroid haloscope, the \textit{Toy CAPPuccino}\cite{12}, is commissioned for data taking at room temperature during 2016 (Fig.\ref{fig-4}). With a minor/major radius of 2/4cm, it is a precursor to the operation of an 80lt cavity at 12T and cryogenic temperatures. The latter will probe the mass spectrum around $1.3-1.8$GHz, with sensitivity in the $g_{a\gamma\gamma}$ values relevant for cold dark matter. \\
An international project led by CAPP is under way for, eventually, a cavity of 50/200cm radius operating at 5T, which will probe the $190-850$MHz mass range. \\
\\
\textbf{Rectangular geometry}\\
In addition to the local haloscopes, a rectangular cavity has recently been included in the CAST helioscope\cite{14} at CERN, forming the \textit{CAST-CAPP/IBS Experiment}\cite{15}, and is currently in commissioning phase. \\
CAST is the leading helioscope experiment since 2003, using a 9T LHC magnet to induce the reverse Primakoff conversion. The RF cavity, housed in one of the magnet bores, forms a new addition to its existing detection techniques. The projected axion mass coverage will be $20-24\mu$eV and, although corresponding to a small part of the mass range, analyzed data are expected within the next couple of years.

\section{Summary}
\label{summary}
A series of direct axion detection searches at the Center for Axion and Precision Physics Research will form part of answering two of the most prominent questions in particle physics: The strong CP problem and the composition of dark matter. The experiments, some in commissioning phase and several under R\&D, will employ cutting-edge technologies improving on all aspects of haloscope operations. \\
These proceedings presented an introduction to the principles of haloscopes, the CAPP R\&D plans and the status of the programme. The Center's goal over the next few years is to cover the major part of the mass and coupling values of dark matter axions, either placing strong bounds on the theoretical expectations or, if the axion exists, discovering it.\\

\section{Acknowledgments}
\label{acknowledgments}
This work was supported by grant IBS-R017-D1-2016-a00 of the Republic of Korea.


\begin{thebibliography}{}

\bibitem{1}
R. D. Peccei and H. Quinn, Phys. Rev. Lett. \textbf{38}, 1440 (1977) 

\bibitem{2}
S. Weinberg, Phys. Rev. Lett. \textbf{40}, 223 (1978)\par
\setlength{\parindent}{.2cm}F. Wilczek, Phys. Rev. Lett. \textbf{40}, 279 (1978) 

\bibitem{3}
J. E. Kim, Phys. Rev. Lett. \textbf{43}, 103 (1979)\par
\setlength{\parindent}{.2cm}M. A. Shifman, A. I. Vainshtein, and V. I. Zakharov, Nucl. Phys. B \textbf{166}, 493 (1980)\par
\setlength{\parindent}{.2cm}A. R. Zhitnitskii, Sov. J. Nucl. Phys. \textbf{31}, 260 (1980)\par
\setlength{\parindent}{.2cm}M. Dine, W. Fischler, and M. Srednicki, Phys. Lett. B \textbf{140}, 199 (1981)

\bibitem{4}
R. Bradley et al, Rev of Mod. Phys. \textbf{75}, 777 (2003) 

\bibitem{5}
http://capp.ibs.re.kr/html/capp\_en/ 

\bibitem{6}
P. Sikivie, Phys. Rev. Lett. \textbf{51}, 1415 (1983)

\bibitem{7}
W. U. Wuensch et al, Phys. Rev. D \textbf{40}, 3153 (1989) 

\bibitem{9}
R. Gupta et al, IEEE Trans. Appl. Supercond. \textbf{26} (2016) 

\bibitem{11}
S. W. Youn, 12th Patras Workshop: \\
https://indico.desy.de/getFile.py/access?contribId=46\&resId=0\&materialId=slides\&confId=13889

\bibitem{12}
 B. R. Ko, 12th Patras Workshop: \par
\setlength{\parindent}{.3cm}https://arxiv.org/abs/1609.03752 

\bibitem{13}
W. H. Chung, 12th Patras Workshop: \\
https://indico.desy.de/getFile.py/access?contribId=50\&resId=0\&materialId=slides\&confId=13889

\bibitem{14}
http://cast.web.cern.ch/CAST/

\bibitem{15}
L. Micelli, 12th Patras Workshop: \\
https://indico.desy.de/getFile.py/access?contribId=53\&resId=0\&materialId=slides\&confId=13889

\end{thebibliography}
\end{document}